\documentclass[prd,aps,twocolumn,showpacs,preprintnumbers,amsmath,amssymb,no footinbib]{revtex4}
\usepackage[dvips]{graphicx} 
\usepackage{amsmath} 
\usepackage{amssymb} 
\usepackage{graphicx}
\usepackage{dcolumn}
\usepackage{bm}
\def\be{\begin{equation}} 
\def\ee{\end{equation}} 
\def\bea{\begin{eqnarray}} 
\def\eea{\end{eqnarray}} 
 
\begin{document} 
 
 
\date{\today} 
 
\title{A Signature of Cosmic Strings Wakes in the CMB Polarization} 
 
\author{Rebecca J. Danos, Robert H. Brandenberger and Gil Holder 
\email[email: ]{rjdanos,,rhb,holder@physics.mcgill.ca}}
 
\affiliation{Department of Physics, McGill University, 
Montr\'eal, QC, H3A 2T8, Canada}

\pacs{98.80.Cq} 
 
\begin{abstract} 

We calculate a signature of cosmic strings in the polarization
of the cosmic microwave background (CMB).
We find that ionization in the wakes behind
moving strings gives rise to extra polarization in a set of rectangular
patches in the sky whose length distribution is scale-invariant. The
length of an individual patch is set by the co-moving Hubble radius at
the time the string is perturbing the CMB. The 
polarization signal is largest for string wakes produced at
the earliest post-recombination time, and for an alignment in which
the photons cross the wake close to the time the wake is created.
The maximal amplitude of the polarization relative to the
temperature quadrupole is set by the overdensity of free
electrons inside a wake which depends on the ionization 
fraction $f$ inside the wake. The signal can be as high as 
$0.06 {\rm \mu K}$ in degree scale polarization for a string
at high redshift (near recombination) and a 
string tension $\mu$ given by $G \mu = 10^{-7}$.

\end{abstract} 
 
\maketitle

\newcommand{\eq}[2]{\begin{equation}\label{#1}{#2}\end{equation}} 
 
\section{Introduction} 

In recent years there has been renewed interest in the possibility that
cosmic strings contribute to the power spectrum of curvature
fluctuations which give rise to the large scale structure and cosmic
microwave background (CMB) anisotropies which we see today.
One of the reasons is that many inflationary models constructed in the 
context of supergravity models lead to the formation of gauge theory
cosmic strings at the end of 
the inflationary phase \cite{Rachel}. Secondly, in a large class of brane
inflation models the formation of cosmic superstrings \cite{Witten}
at the end of inflation is
generic \cite{CS-BI}, and in some cases  (see \cite{Pol1}) these
strings are stable (see also \cite{recentCS} for reviews on fundamental
cosmic strings). Cosmic superstrings are also a possible remnant
of an early Hagedorn phase of string gas cosmology \cite{SGrev}

In models which admit stable strings or superstrings, a scaling solution
of such strings inevitably \cite{Kibble} results as a consequence of cosmological
dynamics (see e.g. \cite{CSrevs} for reviews on cosmic
strings and structure formation).  In a scaling solution, the network
of strings looks statistically the same at any time $t$ if lengths are
scaled to the Hubble radius at that time. The distribution of
strings is dominated by a network of ``infinite" strings 
\footnote{Any string with a mean curvature radius comparable or greater than
the Hubble radius or which extends beyond the Hubble radius is
called ``infinite" or ``long".} with mean
curvature radius and separation being of the order of the Hubble radius.
The scaling solution of infinite strings is maintained by the production
of string loops due to the interaction of long strings. This
leads to a distribution of string loops with a well defined spectrum
(see e.g. \cite{TuBr}) for all radii $R$ smaller than a cutoff radius
set by the Hubble length. Whereas the scaling distribution of the
infinite strings is reasonably well known as a result of detailed
numerical simulations of cosmic string evolution (see
\cite{CSsimuls} for some references), there is still substantial
uncertainty concerning the distribution of string loops. It is,
however, quite clear that the long strings dominate the energy
density of strings. The distinctive signals of strings which we will
focus on are due to the long strings.

Cosmic strings give rise to distinctive signatures in both the CMB
and in the large-scale structure. These signatures are a
consequence of the specific geometry of space produced
by strings. As studied initially in \cite{Vil}, space perpendicular
to a long straight string is locally flat but globally looks like
a cone whose tip coincides with the location of the string 
(the smoothing out of the cone as a consequence
of the internal structure of the cosmic string was worked out
in \cite{Ruth}). The deficit angle is given by
\be \label{deficit}
\alpha \, = \, 8 \pi G \mu \, .
\ee
where $\mu$ is the string tension and $G$ is Newton's constant. 
Hence, a cosmic string moving with velocity
$v$ in the plane perpendicular to its tangent vector will lead
to line discontinuities in the CMB temperature of photons
passing on different sides of the string. The magnitude
of the temperature jump is \cite{KS}
\be \label{KSsig}
{{\delta T} \over T} \, = \, 8 \pi \gamma(v) v G \mu \, ,
\ee
where $\gamma$ is the relativistic gamma factor associated
with the velocity $v$. 

As a consequence of the deficit angle (\ref{deficit}), a moving string
will generate a cosmic string wake, a wedge-shaped region behind
the string (from the point of view of its velocity), a region with twice
the background density \cite{wake} (see Figure 1). Causality
(see e.g. \cite{Traschen}) limits the depth of the distortion of space
due to a cosmic string. The details were worked out in \cite{Joao}
where it was shown that the deficit angle goes to zero quite
rapidly a distance $t$ from the string. Hence, the depth of the
string wake is given by the same length. String wakes lead to
distinctive signatures in the topology of the large-scale
structure, signatures which were explored e.g. in \cite{topology}.

\begin{figure}
\includegraphics[height=5cm]{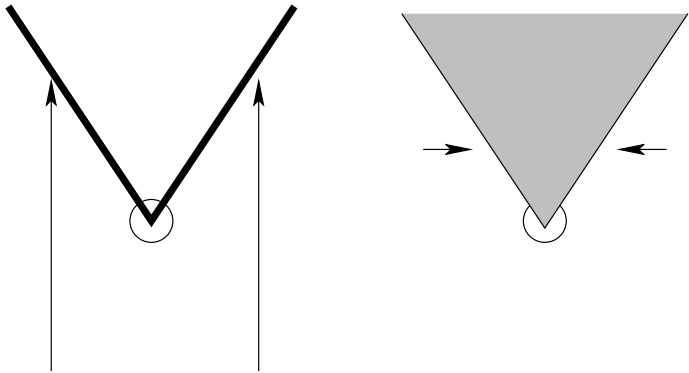}
\caption{Sketch of the mechanism by which a wake behind a moving
string is generated.  Consider a string perpendicular to the plane of
the graph moving straight downward. From the point of view of the
frame in which the string is at rest, matter is moving upwards, as
indicated with the arrows in the left panel. From the point of view
of an observer sitting behind the string (relative to the string motion)
matter flowing past the string receives a velocity kick towards the
plane determined by the direction of the string and the velocity
vector (right panel). This velocity kick towards the plane leads
to a wedge-shaped region behind the string with twice the
background density (the shaded region in the right panel).} \label{fig:1}
\end{figure}

Wakes formed at arbitrarily early times are non-linear density
perturbations. For wakes formed by strings present at times
$t_i > t_{rec}$, where $t_{rec}$ is the time of recombination,
the baryonic matter inside the wake undergoes shocks \cite{Rees}
(see e.g. \cite{Sornborger} for a detailed study). The shocks,
in turn, can ionize the gas - although as it turns out the
residual ionization from decoupling is larger.  
Photons passing through these
ionized regions on their way from the last scattering surface
to the observer can thus be polarized - and it is this
polarization signature which we aim to study here.

The tightest constraints on the contribution of scaling strings
to structure formation (and thus the tightest upper bound
on the tension $\mu$ of the strings) comes from the analysis
of the angular power spectrum of CMB anisotropies. As
discussed in \cite{Perivo,Albrecht,Turok1}, the angular power
spectrum does not have the acoustic ringing which inflation-seeded
perturbations generate. The reason is that the string network is
continuously seeding the growing mode of the curvature
fluctuation variable on super-Hubble scales. Hence, the
fluctuations are ``incoherent" and ``active" as opposed to
``coherent" and ``passive" as in the case of inflation-generated
fluctuations. The contribution of cosmic strings to the
primordial power spectrum of cosmological perturbations
is thus bounded from above, thus leading to an upper bound
\cite{Wyman1,Fraisse,Slosar1,Bevis,Battye} on the string tension of about 
$G \mu < 3 \times 10^{-7}$.  

Past work on CMB temperature maps has shown that
signatures of cosmic strings are easier to identify
in position space than in Fourier space 
\cite{KS, Moessner, Lo, Smoot} and recent
studies show that high angular resolution surveys
such as the South  Pole Telescope project \cite{SPT}
have the potential of improving the limits on the string
tension by an order of magnitude \cite{Amsel, Stewart, Rebecca}. 

To date there has been little work on CMB polarization due
to strings. Most of the existing work focuses on the angular
power spectra of the polarization. Based
on a formalism \cite{Turok1} (see also \cite{Andy})
to include cosmic defects as source terms
in the Boltzmann equations used in CMB codes, the power
spectra of temperature and temperature polarization maps
were worked out \cite{Turok2} in the case of models with
global defects such as global cosmic strings. 
Since in cosmic defect models vector
and tensor modes are as important as the scalar metric
fluctuations \cite{Turok1,Durrer1}, 
a significant B-mode polarization is induced. In
fact, in the case of global strings with a tension close to
the upper bound mentioned above, whereas the amplitude 
of the temperature and E-mode
polarization power spectra are so small as to make the
string signal invisible compared to the signal from the
scale-invariant spectrum of adiabatic fluctuations (e.g.
produced in inflation), the contribution of strings
dominates the amplitude of the B-mode
polarization power spectrum. In the case of local
strings, these conclusions were confirmed in the
more recent analyses of  \cite{Wyman1,Slosar2,Battye,Bevis,Wyman2}. The
maximal amplitude of the B-mode polarization power spectrum
for strings with $G \mu = 3 \times 10^{-7}$ was shown to
be taken on at angular harmonic values of $l \sim 500$
and to be of the order $0.3 \mu K^2$ \cite{Wyman2}.
However, the analyses of \cite{Wyman1,Slosar2,Battye,Bevis,Wyman2} do not
take into account the effects of the gravitational accretion
onto cosmic string wakes. In related work, the conversion of
E-mode to B-mode polarization via the gravitational lensing
induced by cosmic strings was studied in \cite{Francis} 
\footnote{While this paper was being finalized for submission,
a preprint appeared \cite{Durrer2} computing the local B-mode
polarization power spectrum from cosmic strings.}.

Similarly to what was found in the analysis of CMB temperature
maps from cosmic strings, we expect that a position-space
analysis will be more powerful at revealing the key
non-Gaussian signatures of strings in CMB polarization.
Hence, in this work we derive the position space signature
of a cosmic string wake in CMB polarization maps.

In the following section we briefly review cosmic string wakes.
In Section 3 we then analyze the polarization signature of
wakes, and we conclude with some discussion.

\section{Cosmic String Wakes}

Since the strings are relativistic, they generally move with a
velocity of the order of the speed of light. There will be
frequent intersections of strings. The long strings will  chop off
loops, and this leads to the conclusion that the string distribution
will be statistically independent on time scales larger than the
Hubble radius 
\footnote{Hence, to model the effects of strings we will
make use of a toy model introduced in \cite{Periv} and used
in most analytical work on cosmic strings and structure formation
since then: we divide the time interval between recombination 
and the current time $t_0$ into Hubble expansion time steps.
In each time step, there is a distribution of straight
string segments moving in randomly chosen directions with
velocities chosen at random between $0$ and $1$ (in units
of the speed of light). The centers and directions of these
string segments are random, and the string
density corresponds to $N$ strings per Hubble volume,
where $N$ is an integer which is of the order $1$ according
to the scaling of the string network. The distribution of
string segments is uncorrelated at different Hubble times.}.

In this work, we place one string of length $c_1 t_i$ 
\footnote{The constant $c_1$ is of the order $1$ and depends
on the correlation length of the string network as a function 
of the Hubble radius and must be determined from numerical 
simulations of cosmic string evolution.} at a specified time $t_i$
and assume it is moving in transverse direction with a
velocity $v_s$. This string segment will generate a wake, 
and it is the signal of one of these wakes in the CMB 
polarization which we will study in the following.
 
A string segment laid down at time $t_i$.
 will generate a wake 
whose dimensions at that time are the following:
\be \label{size}
c_1 t_i \, \times \, t_i v_s \gamma_s\, \times \, 4 \pi G \mu t_i v_s \gamma_s\, ,
\ee
In the above, the first dimension is the length 
in direction of the string, the second is the depth which depends on
$v_s$ ($\gamma_s$ is the
associated relativistic $\gamma$ factor) , and the third is
the average thickness.

Once the wake is formed, its planar dimensions will expand as the
universe grows in size, and the thickness (defined as the region
of non-linear density) will grow by gravitational accretion. The
accretion of matter onto a cosmic string wake
was studied in \cite{Albert1,Leandros1} in the case of the dark matter 
being cold, and in \cite{Leandros1,Leandros2} in the case of the dark 
matter being hot 
\footnote{If the strings have lots of small-scale structure then
they will have an effective tension which is less than the effective
energy density \cite{wiggly}. This will lead to a local gravitational
attraction of matter towards the string, a smaller transverse
velocity, and hence to string filaments instead of wakes. The
gravitational accretion onto string filaments was studied in
\cite{Aguirre}.}. We are interested in the case of cold dark matter.

We consider mass planes at a fixed initial
comoving distance $q$ above the center of the wake. The
corresponding physical height is
\be
h(q, t) \, = \, a(t) \bigl[ q - \psi(q, t) \bigr] \, ,
\ee
where $\psi(q,t)$ is the comoving displacement induced by the
gravitational accretion onto the wake. For cold dark matter,
the initial conditions for $\psi(q)$ are $\psi(q, t_i) = \dot{\psi(q, t_i)} = 0$.
The goal of the analysis is to find the
thickness of the wake at all times $t > t_i$. The thickness is
defined as the physical height $h$ above the center of
the wake of the matter shell which is beginning to fall
towards the wake, i.e. for which $ \dot{h(q,t)} = 0$.
In the Zel'dovich approximation, we first consider
the equation of motion for $h$ 
obtained by treating the source (the initial surface density $\sigma$ 
of the wake) in the Newtonian limit, i.e.
\be
\ddot{h} \, = \ - \frac{\partial \Phi}{\partial h} \, ,
\ee
where $\Phi$ is the Newtonian gravitational potential given
by the Poisson equation
\be
\frac{\partial^2 \Phi}{\partial^2 h} \, = \, 4 \pi G \bigl[ \rho + \sigma \delta(h) \bigr]
\ee
($\rho(t)$ being the background energy density), 
and then by linearizing the resulting equation
in $\psi$. The mean surface density  is
\be
\sigma(t) \, = \, 4 \pi G \mu t_i v_s \gamma_s \bigl( \frac{t}{t_i} \bigr)^{2/3} \rho(t) \, .
\ee

The result of the computation of the value of the comoving displacement $q_{nl}$ which
is ``turning around" at the time $t$ for a wake laid down at time $t_i$ is \cite{Leandros1}
\be
q_{nl}(t, t_i) \, = \, \psi_0  \bigl( \frac{t}{t_i} \bigr)^{2/3} 
\ee
with
\be \label{psieq}
\psi_0(t_i) \, = \, \frac{24 \pi}{5} G \mu v_s \gamma_s (z(t_i) + 1)^{-1/2} t_0 \, .
\ee
This corresponds to a physical height of
\bea
h(t, t_i) \, &=& \, a(t) \bigl[ q_{nl}(t, t_i)  - \psi(q_{nl}, t) \bigr] \, \simeq \, a(t) q_{nl}(t, t_i) \nonumber \\
                 &=& \, \psi_0 \frac{z_i + 1}{(z + 1)^2} \, ,
\eea
where $z_i$ and $z$ are the redshifts corresponding to the times $t_i$ and $t$, respectively.
These formulas agree with what is expected from linear cosmological perturbation theory:
the fractional density perturbation should increase linearly in the scale factor which
means that the comoving width of the wake must grow linearly with $a(t)$.

Let us return to the geometry of the string segment. The tangent vector 
to the string and the direction of motion
of the string determine a two dimensional hypersurface in
space (the ``string plane"). If neither the string tangent vector nor the 
velocity vector have a radial component (radial with respect
to the co-moving point of our observer), then the photons
will cross each point of the string wake at the same time,
and the wake will correspond to a rectangle in the sky whose planar dimensions
are
\be
c_1 t_1 \times t_i v_s \gamma_s \, .
\ee
However, if the normal vector of the string plane is not radial, then one of the
planar dimensions will be reduced by a trigonometric factor ${\rm cos} (\theta)$
depending on the angle $\theta$ between the normal vector and the radial
vector. In addition, photons which we detect today did not pass the wake
at the same time. This will lead to a gradient of the polarization signal across
the projection of the wake onto the CMB sky. To simplify the analysis, in
the following we will assume that $\theta$ is small.

In the following section we will need the expression for the number density 
$n_e(t, t_i)$ of free electrons at time $t$ in a wake which was laid down at the 
time $t_i$. The initial number density is
\be
n_e(t_i, t_i) \, \simeq \, f \rho_B(t_i) m_{p}^{-1} \, ,
\ee
where $f$ is the ionization fraction, $\rho_B$ is the energy density in baryons, and
$m_p$ is the proton mass. Taking into account that for $t > t_i$ the number
density redshifts as the inverse volume we get
\be
n_e(t, t_i) \, \simeq \, f \rho_B(t_i) m_p^{-1} \bigl( \frac{z(t) + 1}{z(t_i) + 1} \bigr)^{3} \, .
\ee
                
\section{Analysis}

If unpolarized cosmic microwave background radiation with a quadrupolar
anisotropy scatters off free electrons, then the scattered radiation is
polarized (see e.g. \cite{polariz} for reviews of the theory of CMB polarization).
The magnitude of the polarization depends on the Thomson cross section
$\sigma_T$ and on the integral of the number density of electrons along the
null geodesic of radiation. Our starting formula is
\be \label{polar1}
P({\bf{n}}) \, \simeq \, \frac{1}{10} \bigl( \frac{3}{4 \pi} \bigr)^{1/2} \tau_T Q \, ,
\ee
where $P({\bf{n}}) $ is the magnitude of polarization of radiation from the
direction $\bf{n}$, $Q$ is the temperature quadrupole, and
\be \label{integral}
\tau_T \, = \, \sigma_T \int n_e(\chi) d\chi \, ,
\ee
where the integral is along the null geodesic in terms of conformal time.

Let us now estimate the contribution to the polarization amplitude from 
CMB radiaton passing through a single wake at time $t$, assuming that
the wake was laid down at the time $t_i$ and has thus had time to
grow from time $t_i$ to the time $t$ when the photons are
crossing it. We assume that the photons cross in perpendicular direction.
Since the wake is thin, we can estimate the integral in Eq. (\ref{integral})
by
\be \label{integral2}
\tau_T \, \sim \, 2 \sigma_T n_e(t, t_i) \bigl( z(t) + 1 \bigr) h(t, t_i) \, 
\ee
where the factor $2$ is due to the fact that the width of the wake is
twice the hight, and the redshift factor comes from the Jacobean 
transformation between $\chi$ and position. Inserting Eq. (\ref{integral2})
into the expression (\ref{polar1}) for the polarization amplitude
and using the value for the height and the number density of
electrons obtained in the previous section we find
\bea
\frac{P}{Q} \, &\simeq& \, \frac{1}{5} \bigl( \frac{3}{4 \pi} \bigr)^{1/2} \sigma_T
f \rho_B(t_i) m_p^{-1} \nonumber \\
& & \times \frac{(z(t) + 1)^2}{(z(t_i) + 1)^2} \psi_0(t_i) \, .
\eea
Inserting the formula for $\psi_0(t_i)$ from Eq. (\ref{psieq}), expressing the
baryon density at $t_i$ in terms of the current baryon density, and the
latter in terms of the baryon fraction $\Omega_B$ and the total energy 
density $\rho_c$ at the current time $t_0$, we obtain
\bea \label{result}
\frac{P}{Q} \, &\simeq& \, \frac{24 \pi}{25}  \bigl( \frac{3}{4 \pi} \bigr)^{1/2} \sigma_T f G \mu
v_s \gamma_s \nonumber \\
& & \times \Omega_B \rho_c(t_0) m_p^{-1} t_0 \bigl( z(t) + 1 \bigr)^2 \bigl( z(t_i) + 1 \bigr)^{1/2} \, .
\eea

{F}rom Equation (\ref{result}) we see that the polarization signal is larger for wakes laid
down early, i.e. close to the time of recombination. For fixed $t_i$, the signal is largest
for configurations where the photons we observe today cross the wake at the earliest
possible time, i.e. for the largest $z(t)$ (obviously, $t$ is constrained to be larger
than $t_i$, otherwise our formula for the height is not applicable). To get an order
of magnitude estimate of the magnitude of the polarization signal, we take 
$z(t_i) \sim z(t) \sim 10^3$. Inserting the value of the Thomson cross section, the
proton mass and the current time we get
\be
\frac{P}{Q} \, \sim \, f G \mu v_s \gamma_s \Omega_B 
\bigl( \frac{z(t) + 1}{10^3} \bigr)^2 \bigl( \frac{z(t_i) + 1}{10^3} \bigr)^3
10^7 \, .
\ee

The ionization fraction of baryonic matter drops off after recombination,
but it does not go to zero. As already discussed in \cite{Zel,Peebles} 
there is remnant residual ionization of the matter. As computed in
\cite{Peebles} (see also \cite{Gil}), the residual ionization fraction 
$f$ tends to a limiting
value of between $10^{-5}$ and $10^{-4}$ at late times after recombination.

Shocks inside the wake will lead to extra ionization. However, the
resulting contribution to the ionization fraction is negligible for
the range of string tensions we are interested in. To see this, we
follow the analysis of \cite{Rees}.  A particle streaming towards 
a cosmic string wake has velocity
\be
v_i \, = \, 4 \pi G \mu v_s \gamma_s
\ee
and kinetic energy  $\frac{1}{2}mv_i^2$. The particles will undergo
shocks and thermalize. Equating the initial kinetic energy density 
with the final thermal energy density
(assuming that the particles are distributed according to an approximate
ideal gas law) yields a temperature inside the wakes of
\be
T \, \simeq \, 7 \times 10^3 \left(\frac{G \mu}{2\times
10^{-6}}\right)^2 (v_s \gamma_s)^2 \, K \, .  
\ee
We then compute the Boltzmann factor to determine the ratio
of ionized Hydrogen to ground state Hydrogen which is the ionization
fraction.  For string tensions of order $10^{-6}$ this results in
ionization fractions of the order $10^{-9}$ which is considerably less
than the residual ionization.  For more realistic string tensions
compatible with current bounds, $10^{-7}$, the ionization fraction due
to shocks is negligible compared to the residual ionization fraction.  At the
temperatures considered for a string tension of $10^{-7}$ at lower
redshifts there would be star formation and hence an ionization
fraction due to that effect  since the wake would contain molecular
hydrogen and satisfy the appropriate conditions.  However, this is not
an issue at redshifts under consideration.
 
Note that even though the extra ionization inside the wake is negligible
compared to the overall ionization level, the wake is a locus of extra 
energy density. Thus, even if the ionization fraction is homogeneous in space, the
inhomogeneous distribution of matter will lead to a specific polarization
signature.

The position space signal of the polarization produced by a wake will be very
specific: for a wake whose normal vector is in radial direction,
it will be a rectangle in the sky with angular dimensions
corresponding to the comoving size
\be \label{size2}
c_1 t_i \bigl( z(t_i) + 1 \bigr) \times v_s \gamma_s t_i \bigl( z(t_i) + 1 \bigr) 
\ee
(see Eq. (\ref{size})). If the angle of the normal of the string plane to the radial
normal vector is $\theta \neq 0$, then one of the planar dimensions is
reduced by a factor of ${\rm cos} (\theta)$. However, the light travel time
through the wake is increased by a factor of $[{\rm cos} (\theta)]^{-1}$.
Thus, the wake becomes a bit smaller but the signal strength increases.
The average amplitude of the polarization is given 
by Eq. (\ref{result}) - a value obtained using the average thickness of the
wake. However, since the thickness of the wake increases linearly
in direction of the string motion, the amplitude of the polarization 
signal will also increase linearly in this direction. A second source
for the linear increase in the amplitude is the fact that
$z(t_i)$ is increasing as we go away from the tip of the cone (by
an amount corresponding to one Hubble expansion time when comparing
the tip of the cone to the end). The direction of
the polarization vector depends on the relative orientation between
the string and the CMB quadrupole. In Figure 2 we give a sketch
of the signal. The amplitude of the polarization is proportional
to the length of the arrow, the direction of the polarization is given
by the direction of the arrow.
Since the direction of the variation of the polarization strength is determined
by the string and is therefore uncorrelated with the direction of the CMB
quadrupole, on the average the E-mode and B-mode strengths of
the polarization signal will be the same.

\begin{figure}
\includegraphics[height=6cm]{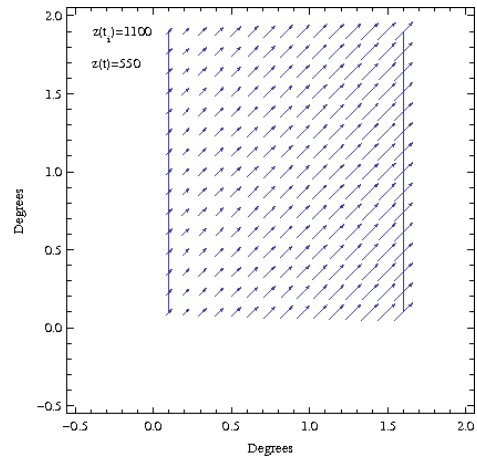}
\caption{The polarization signal of a single wake which is taken to be
perpendicular to the line of sight between us and the center of the string segment.
The tip of the wake (position of the string at the time the wake is laid down) is
the vertical edge of the rectangle, and the string velocity vector is pointing 
horizontally to the left The length and direction of the arrows indicate the magnitude
and orientation of the polarization vector. We have assumed that the
variation of the quadrupole vector across the plane of the wake is negligible. 
Note that the angle between the velocity vector of the string and the CMB
quadrupole is random. We
have assumed that the quadrupole vector is at an angle relative to the plane of
the wake. This determines the direction of the polarization vector we have
drawn. The orientation we have chosen in this figure corresponds to an
almost pure B-mode.}   \label{fig:2}
\end{figure}

Let us now discuss the magnitude of our effect. We will consider
the value $G \mu = 3 \times 10^{-7}$ which is the current upper
bound on the string tension \cite{Wyman1} (using the assumptions
about the cosmic string scaling solution made in these papers).
Wakes produced close to the time of recombination inherit the
ionization fraction of the universe at that time. Taking a value
of $f = 10^{-3}$ (which is smaller than the ionization fraction
until redshift $z \simeq 600$ \cite{Gil}), we obtain a polarization
amplitude of $P \sim 10^{-2}$ $\mu {\rm K}$ which is larger than
the background in the B-mode polarization arising from weak
lensing of the primordial perturbations \cite{Hirata,Holder2} for
l-values of about $100$.

It is instructive to compare our polarization signal from a cosmic
string wake with the expected noise due to the Gaussian
fluctuations. In Fig. 3 we have superimposed the map
of the Q-mode polarization from a cosmic string wake laid
down during the first Hubble time after recombination
with a corresponding Q-mode map due to Gaussian noise
of the concordance $\Lambda$CDM model.
The string parameters are the same as mentioned in the
previous paragraph. We chose the orientation of the
string relative to the CMB quadrupole such that the power
in the Q-mode is half the total power. As the value of the 
CMB quadrupole we used $30 {\rm{\mu K}}$. To render
the string signal visible in the Q-mode map (in which the
noise is much larger than in a B-mode map) we
multiplied the string signal by $100$. In this case
the string signal is clearly visible be eye. The
brightest edge (the vertical edge on the right side)
corresponds to the position of the string when it
begins to generate the wake, not at the position
at the end of the time interval being considered (the
vertical edge on the left). Note the difference compared
to the Kaiser-Stebbins effect in the CMB temperature
maps: in this case the brightest edge corresponds to
the location of the string when the photons are passing
by it. 

\begin{figure}
\includegraphics[height=6cm]{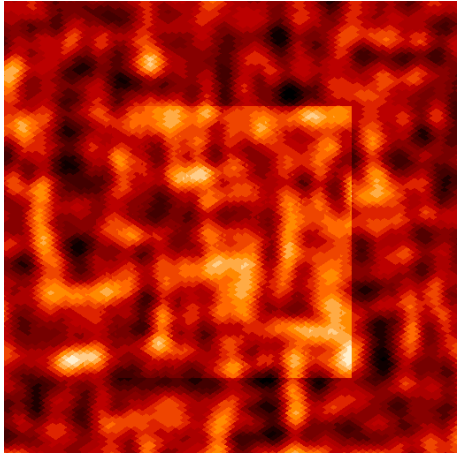}
\caption{The Q-mode polarization signal of a single wake which is taken to be
perpendicular to the line of sight between us and the center of the string segment,
superimposed on the Gaussian noise signal which is expected to
dominate the total power spectrum. The string signal is multiplied by a factor
of $100$ to render it visible by eye.
The tip of the wake (position of the string at the time the wake is laid down) is
the vertical bright edge of the right side, and the string velocity vector is pointing 
horizontally to the left. At the final string position the wake thickness
vanishes and there is no polarization discontinuity line. We have assumed that the
variation of the quadrupole vector across the plane of the wake is negligible
and that the quadrupole vector is at an angle relative to the plane of
the wake such that the Q-mode picks up half of the polarization power.}   
\label{fig:3}
\end{figure}

Without boosting the string signal by a large factor, it would
not be visible by eye. However, the distinctive lines in the map can be
searched for by edge detection algorithms such as the Canny algorithm
which was used to study the string signal in CMB temperature maps.
In the studies of \cite{Rebecca} it was found that the cosmic string lines
in temperature maps can be picked out if the string signal accounts for
less than $0.2 \%$ of the power. Thus, it should be able to easily pick
out the string signal in the Q-mode polarization maps. In B-mode
polarization maps the string signal would be much easier to detect.

\section{Conclusions and Discussion}
 
In this Letter we have discussed a position space signal of a
cosmic string wake in CMB polarization maps. In the same
way that the line discontinuities in the CMB temperature maps
predicted by the Kaiser-Stebbins (KS) effect yield a promising way
to constrain/detect cosmic strings in the CMB (see e.g.
\cite{Moessner,Lo,Smoot,Amsel,Stewart,Rebecca}, we believe that the
signal discussed in this paper will play a similar role once
CMB polarization maps become available. 

We have shown that a single wake will produce a rectangular
patch in the sky of dimensions given by equation (\ref{size2}), average
magnitude given by equation (\ref{result}) and amplitude increasing
linearly in one direction across the patch. For a value of
the string tension $G \mu = 3 \times 10^{-7}$ (the current upper
limit), the amplitude of the signal is within the range of 
planned polarization experiments for wakes
produced sufficiently close to the surface of recombination.
These wakes are also the most numerous ones. 
The brightest edge in the polarization map corresponds to the
beginning location of the string, not the final location.
Since the KS discontinuity in the CMB temperature map 
will occur along the line corresponding to the final position, 
the polarization signal discussed here provides a cross-check
on a possible string interpretation of a KS signal.  

A scaling distribution
of strings will yield a distribution of patches in the sky, the most
numerous ones and the ones with the largest polarization
amplitude being set by wakes laid down at times close to the
time of recombination which are crossed by the CMB photons
at similarly early times. 

\begin{acknowledgments} 
 
This work is supported in part by a NSERC Discovery Grants and 
by funds from the CRC Program to RB and GH, by a Killam Research 
Fellowship awarded to R.B, and by CIfAR (GH). 
We thank Andrew Frey and Oscar Hernandez for useful discussions.

\end{acknowledgments}


\begin{thebibliography}{99} 
 
 \bibitem{Rachel}
R.~Jeannerot,
  ``A Supersymmetric SO(10) Model with Inflation and Cosmic Strings,''
  Phys.\ Rev.\  D {\bf 53}, 5426 (1996)
  [arXiv:hep-ph/9509365];\\
R.~Jeannerot, J.~Rocher and M.~Sakellariadou,
  ``How generic is cosmic string formation in SUSY GUTs,''
  Phys.\ Rev.\  D {\bf 68}, 103514 (2003)
  [arXiv:hep-ph/0308134].

 \bibitem{Witten}
E.~Witten,
  ``Cosmic Superstrings,''
  Phys.\ Lett.\  B {\bf 153}, 243 (1985).

 \bibitem{CS-BI}
S.~Sarangi and S.~H.~H.~Tye,
  ``Cosmic string production towards the end of brane inflation,''
  Phys.\ Lett.\  B {\bf 536}, 185 (2002)
  [arXiv:hep-th/0204074].

\bibitem{Pol1}
E.~J.~Copeland, R.~C.~Myers and J.~Polchinski,
  ``Cosmic F- and D-strings,''
  JHEP {\bf 0406}, 013 (2004)
  [arXiv:hep-th/0312067].

 \bibitem{recentCS}
 A.~C.~Davis and T.~W.~B.~Kibble,
  ``Fundamental cosmic strings,''
  Contemp.\ Phys.\  {\bf 46}, 313 (2005)
  [arXiv:hep-th/0505050];\\
  M.~Sakellariadou,
  ``Cosmic Superstrings,''
  arXiv:0802.3379 [hep-th].
    
 \bibitem{SGrev}
 R.~H.~Brandenberger and C.~Vafa, 
  ``Superstrings In The Early Universe,'' 
  Nucl.\ Phys.\ B {\bf 316}, 391 (1989).;\\
A.~Nayeri, R.~H.~Brandenberger and C.~Vafa, 
  ``Producing a scale-invariant spectrum of perturbations in a Hagedorn phase 
  of string cosmology,''
 Phys.\ Rev.\ Lett.\  {\bf 97}, 021302 (2006)   [arXiv:hep-th/0511140];\\
 R.~H.~Brandenberger, A.~Nayeri, S.~P.~Patil and C.~Vafa,
  ``String gas cosmology and structure formation,''
  arXiv:hep-th/0608121;\\
  R.~H.~Brandenberger,
  ``String Gas Cosmology,''
  arXiv:0808.0746 [hep-th].
   
 \bibitem{Kibble}
 T.~W.~B.~Kibble,
  ``Phase Transitions In The Early Universe,''
  Acta Phys.\ Polon.\  B {\bf 13}, 723 (1982);\\
  T.~W.~B.~Kibble,
  ``Some Implications Of A Cosmological Phase Transition,''
  Phys.\ Rept.\  {\bf 67}, 183 (1980).
  
\bibitem{CSrevs}
A. Vilenkin and E.P.S. Shellard;
\textit{Cosmic Strings and Other Topological Defects},
(Cambridge Univ. Press, Cambridge, 1994);\\
M.~B.~Hindmarsh and T.~W.~Kibble,
``Cosmic strings,''
Rept.\ Prog.\ Phys.\  {\bf 58}, 477 (1995)
[arXiv:hep-ph/9411342];\\
R.~H.~Brandenberger,
``Topological defects and structure formation,''
Int.\ J.\ Mod.\ Phys.\ A {\bf 9}, 2117 (1994)
[arXiv:astro-ph/9310041].
  
 \bibitem{TuBr}
N.~Turok and R.~H.~Brandenberger,
  ``Cosmic Strings And The Formation Of Galaxies And Clusters Of Galaxies,''
  Phys.\ Rev.\ D {\bf 33}, 2175 (1986);\\
H. Sato, ``Galaxy Formation by Cosmic Strings,''
  Prog. Theor. Phys.\  {\bf 75}, 1342 (1986);\\
A. Stebbins, ``Cosmic Strings and Cold Matter'',
  Ap. J. (Lett.) {\bf 303}, L21 (1986).
 
\bibitem{CSsimuls}
A.~Albrecht and N.~Turok,
  ``Evolution Of Cosmic Strings,''
  Phys.\ Rev.\ Lett.\  {\bf 54}, 1868 (1985);\\
D.~P.~Bennett and F.~R.~Bouchet,
  ``Evidence For A Scaling Solution In Cosmic String Evolution,''
  Phys.\ Rev.\ Lett.\  {\bf 60}, 257 (1988);\\
B.~Allen and E.~P.~S.~Shellard,
  ``Cosmic String Evolution: A Numerical Simulation,''
  Phys.\ Rev.\ Lett.\  {\bf 64}, 119 (1990);\\
C.~Ringeval, M.~Sakellariadou and F.~Bouchet,
  ``Cosmological evolution of cosmic string loops,''
  JCAP {\bf 0702}, 023 (2007)
  [arXiv:astro-ph/0511646];\\
V.~Vanchurin, K.~D.~Olum and A.~Vilenkin,
  ``Scaling of cosmic string loops,''
  Phys.\ Rev.\  D {\bf 74}, 063527 (2006)
  [arXiv:gr-qc/0511159].
  
\bibitem{Vil}
A.~Vilenkin,
  ``Gravitational Field Of Vacuum Domain Walls And Strings,''
  Phys.\ Rev.\  D {\bf 23}, 852 (1981).

\bibitem{Ruth}
R.~Gregory,
  ``Gravitational Stability of Local Strings,''
  Phys.\ Rev.\ Lett.\  {\bf 59}, 740 (1987).
  
\bibitem{KS}
N.~Kaiser and A.~Stebbins,
  ``Microwave Anisotropy Due To Cosmic Strings,''
  Nature {\bf 310}, 391 (1984).

\bibitem{wake}
J.~Silk and A.~Vilenkin,
  ``Cosmic Strings And Galaxy Formation,''
  Phys.\ Rev.\ Lett.\  {\bf 53}, 1700 (1984).
  
\bibitem{Traschen}
J.~H.~Traschen,
  ``Causal Cosmological Perturbations And Implications For The Sachs-Wolfe
  Effect,''
  Phys.\ Rev.\  D {\bf 29}, 1563 (1984);\\
  J.~H.~Traschen,
  ``Constraints On Stress Energy Perturbations In General Relativity,''
  Phys.\ Rev.\  D {\bf 31}, 283 (1985).
  
\bibitem{Joao}
J.~C.~R.~Magueijo,
  ``Inborn metric of cosmic strings,''
  Phys.\ Rev.\  D {\bf 46}, 1368 (1992).
 
\bibitem{topology}
 R.~H.~Brandenberger, D.~M.~Kaplan and S.~A.~Ramsey,
  ``Some statistics for measuring large scale structure,''
  arXiv:astro-ph/9310004;\\
  D.~Mitsouras, R.~H.~Brandenberger and P.~Hickson,
  ``Topological Statistics and the LMT Galaxy Redshift Survey,''
  arXiv:astro-ph/9806360.

\bibitem{Rees}
 M. Rees,
 ``Baryon concentrations in string wakes at $z \geq 200$:
 implications for galaxy formation and large-scale structure,"
 Mon. Not. R. astr. Soc. {\bf{222}}, 27p (1986).
  
 \bibitem{Sornborger}
A.~Sornborger, R.~H.~Brandenberger, B.~Fryxell and K.~Olson,
  ``The structure of cosmic string wakes,''
  Astrophys.\ J.\  {\bf 482}, 22 (1997)
  [arXiv:astro-ph/9608020].

\bibitem{Perivo}
L.~Perivolaropoulos,
  ``Spectral Analysis Of Microwave Background Perturbations Induced By Cosmic
  Strings,''
  Astrophys.\ J.\  {\bf 451}, 429 (1995)
  [arXiv:astro-ph/9402024].

\bibitem{Albrecht}
J.~Magueijo, A.~Albrecht, D.~Coulson and P.~Ferreira,
  ``Doppler peaks from active perturbations,''
  Phys.\ Rev.\ Lett.\  {\bf 76}, 2617 (1996)
  [arXiv:astro-ph/9511042].

\bibitem{Turok1}
U.~L.~Pen, U.~Seljak and N.~Turok,
  ``Power spectra in global defect theories of cosmic structure formation,''
  Phys.\ Rev.\ Lett.\  {\bf 79}, 1611 (1997)
  [arXiv:astro-ph/9704165].

\bibitem{Wyman1}
L.~Pogosian, S.~H.~H.~Tye, I.~Wasserman and M.~Wyman,
  ``Observational constraints on cosmic string production during brane
  inflation,''
  Phys.\ Rev.\  D {\bf 68}, 023506 (2003)
  [Erratum-ibid.\  D {\bf 73}, 089904 (2006)]
  [arXiv:hep-th/0304188];\\
M.~Wyman, L.~Pogosian and I.~Wasserman,
  ``Bounds on cosmic strings from WMAP and SDSS,''
  Phys.\ Rev.\  D {\bf 72}, 023513 (2005)
  [Erratum-ibid.\  D {\bf 73}, 089905 (2006)]
  [arXiv:astro-ph/0503364].

\bibitem{Fraisse}
A.~A.~Fraisse,
  ``Limits on Defects Formation and Hybrid Inflationary Models with
  Three-Year WMAP Observations,''
  JCAP {\bf 0703}, 008 (2007)
  [arXiv:astro-ph/0603589].
  
 \bibitem{Slosar1} 
U.~Seljak, A.~Slosar and P.~McDonald,
  ``Cosmological parameters from combining the Lyman-alpha forest with CMB,
  galaxy clustering and SN constraints,''
  JCAP {\bf 0610}, 014 (2006)
  [arXiv:astro-ph/0604335].
  
 \bibitem{Bevis}
N.~Bevis, M.~Hindmarsh, M.~Kunz and J.~Urrestilla,
  ``CMB power spectrum contribution from cosmic strings using  field-evolution
  simulations of the Abelian Higgs model,''
  Phys.\ Rev.\  D {\bf 75}, 065015 (2007)
  [arXiv:astro-ph/0605018];\\
N.~Bevis, M.~Hindmarsh, M.~Kunz and J.~Urrestilla,
  ``Fitting CMB data with cosmic strings and inflation,''
Phys.\ Rev.\ Lett.\  {\bf 100}, 021301 (2008)
  [arXiv:astro-ph/0702223].
  
\bibitem{Battye}
R.~A.~Battye, B.~Garbrecht and A.~Moss,
  ``Constraints on supersymmetric models of hybrid inflation,''
  JCAP {\bf 0609}, 007 (2006)
  [arXiv:astro-ph/0607339];\\
R.~A.~Battye, B.~Garbrecht, A.~Moss and H.~Stoica,
  ``Constraints on Brane Inflation and Cosmic Strings,''
  JCAP {\bf 0801}, 020 (2008)
  [arXiv:0710.1541 [astro-ph]].

\bibitem{Moessner}
R.~Moessner, L.~Perivolaropoulos and R.~H.~Brandenberger,
  ``A Cosmic string specific signature on the cosmic microwave background,''
  Astrophys.\ J.\  {\bf 425}, 365 (1994)
  [arXiv:astro-ph/9310001].

\bibitem{Lo} 
A.~S.~Lo and E.~L.~Wright,
  ``Signatures of cosmic strings in the cosmic microwave background,''
  arXiv:astro-ph/0503120.
  
\bibitem{Smoot} 
E.~Jeong and G.~F.~Smoot,
  ``Search for cosmic strings in CMB anisotropies,''
  Astrophys.\ J.\  {\bf 624}, 21 (2005)
  [arXiv:astro-ph/0406432];\\
E.~Jeong and G.~F.~Smoot,
  ``The Validity of the Cosmic String Pattern Search with the Cosmic
  Microwave Background,''
  arXiv:astro-ph/0612706.
  
\bibitem{SPT}
 J.~E.~Ruhl {\it et al.}  [The SPT Collaboration],
  ``The South Pole Telescope,''
  Proc.\ SPIE Int.\ Soc.\ Opt.\ Eng.\  {\bf 5498}, 11 (2004)
  [arXiv:astro-ph/0411122].
 
\bibitem{Amsel}
S.~Amsel, J.~Berger and R.~H.~Brandenberger,
  ``Detecting Cosmic Strings in the CMB with the Canny Algorithm,''
  JCAP {\bf 0804}, 015 (2008)
  [arXiv:0709.0982 [astro-ph]].
  
\bibitem{Stewart}
A.~Stewart and R.~Brandenberger,
  ``Edge Detection, Cosmic Strings and the South Pole Telescope,''
  arXiv:0809.0865 [astro-ph].
  
\bibitem{Rebecca}
R.~J.~Danos and R.~H.~Brandenberger,
  ``Canny Algorithm, Cosmic Strings and the Cosmic Microwave Background,''
  arXiv:0811.2004 [astro-ph];\\
R.~J.~Danos and R.~H.~Brandenberger,
  ``Searching for Signatures of Cosmic Superstrings in the CMB,''
  arXiv:0910.5722 [astro-ph.CO].
  
 \bibitem{Andy}
A.~J.~Albrecht, R.~A.~Battye and J.~Robinson,
  ``A detail study of defect models for cosmic structure formation,''
  Phys.\ Rev.\  D {\bf 59}, 023508 (1999)
  [arXiv:astro-ph/9711121].
   
\bibitem{Turok2}
U.~Seljak, U.~L.~Pen and N.~Turok,
  ``Polarization of the Microwave Background in Defect Models,''
  Phys.\ Rev.\ Lett.\  {\bf 79}, 1615 (1997)
  [arXiv:astro-ph/9704231].

\bibitem{Durrer1}
R.~Durrer, M.~Kunz and A.~Melchiorri,
  ``Cosmic Microwave Background Anisotropies from Scaling Seeds: Global Defect
  Models,''
  Phys.\ Rev.\  D {\bf 59}, 123005 (1999)
  [arXiv:astro-ph/9811174].
  
\bibitem{Slosar2}
 U.~Seljak and A.~Slosar,
  ``B polarization of cosmic microwave background as a tracer of strings,''
  Phys.\ Rev.\  D {\bf 74}, 063523 (2006)
  [arXiv:astro-ph/0604143].
  
\bibitem{Wyman2}
L.~Pogosian, I.~Wasserman and M.~Wyman,
  ``On vector mode contribution to CMB temperature and polarization from  local
  strings,''
  arXiv:astro-ph/0604141;\\
L.~Pogosian and M.~Wyman,
  ``B-modes from Cosmic Strings,''
  Phys.\ Rev.\  D {\bf 77}, 083509 (2008)
  [arXiv:0711.0747 [astro-ph]].

\bibitem{Francis}
K.~Benabed and F.~Bernardeau,
  ``Cosmic string lens effects on CMB polarization patterns,''
  Phys.\ Rev.\  D {\bf 61}, 123510 (2000).

\bibitem{Durrer2}  
J.~Garcia-Bellido, R.~Durrer, E.~Fenu, D.~G.~Figueroa and M.~Kunz,
  ``The local B-polarization of the CMB: a very sensitive probe of cosmic
  defects,''
  arXiv:1003.0299 [astro-ph.CO].

\bibitem{Periv}
L.~Perivolaropoulos,
  ``COBE versus cosmic strings: An Analytical model,''
  Phys.\ Lett.\  B {\bf 298}, 305 (1993)
  [arXiv:hep-ph/9208247];\\
L.~Perivolaropoulos,
  ``Statistics of microwave fluctuations induced by topological defects,''
  Phys.\ Rev.\  D {\bf 48}, 1530 (1993)
  [arXiv:hep-ph/9212228].
  
\bibitem{Albert1}
A.~Stebbins, S.~Veeraraghavan, R.~H.~Brandenberger, J.~Silk and N.~Turok,
  ``Cosmic String Wakes,''
  Astrophys.\ J.\  {\bf 322}, 1 (1987).

\bibitem{Leandros1}
R.~H.~Brandenberger, L.~Perivolaropoulos and A.~Stebbins,
  ``Cosmic Strings, Hot Dark Matter and the Large-Scale Structure of the Universe,"
  Int.\ J.\ Mod.\ Phys.\  A {\bf 5}, 1633 (1990).
  
\bibitem{Leandros2}
L.~Perivolaropoulos, R.~H.~Brandenberger and A.~Stebbins,
  ``Dissipationless Clustering Of Neutrinos In Cosmic String Induced Wakes,''
  Phys.\ Rev.\  D {\bf 41}, 1764 (1990).
  
\bibitem{wiggly}
 B.~Carter,
  ``Integrable equation of state for noisy cosmic string,''
  Phys.\ Rev.\  D {\bf 41}, 3869 (1990);\\
A.~Vilenkin,
   ``Effect of Small Scale Structure on the Dynamics of Cosmic Strings,"
  Phys.\ Rev.\  D {\bf 41}, 3038 (1990).
  
\bibitem{Aguirre}
 A.~N.~Aguirre and R.~H.~Brandenberger,
  ``Accretion of hot dark matter onto slowly moving cosmic strings,''
  Int.\ J.\ Mod.\ Phys.\  D {\bf 4}, 711 (1995)
  [arXiv:astro-ph/9505031].

\bibitem{polariz}
W.~Hu,
  ``Reionization Revisited: Secondary CMB Anisotropies and Polarization,''
  Astrophys.\ J.\  {\bf 529}, 12 (2000)
  [arXiv:astro-ph/9907103];\\
O.~Dore, G.~Holder, M.~Alvarez, I.~T.~Iliev, G.~Mellema, U.~L.~Pen and P.~R.~Shapiro,
  ``The Signature of Patchy Reionization in the Polarization Anisotropy of the
  CMB,''
  Phys.\ Rev.\  D {\bf 76}, 043002 (2007)
  [arXiv:astro-ph/0701784].
  
\bibitem{Zel}
I.~D.~Novikov and Y.~B.~Zeldovic,
  ``Cosmology,''
  Ann.\ Rev.\ Astron.\ Astrophys.\  {\bf 5}, 627 (1967).
  
\bibitem{Peebles}
P.~J.~E.~Peebles,
  ``Recombination Of The Primeval Plasma,''
  Astrophys.\ J.\  {\bf 153}, 1 (1968).

\bibitem{Gil}
M.~Kaplinghat, M.~Chu, Z.~Haiman, G.~Holder, L.~Knox and C.~Skordis,
  ``Probing the Reionization History of the Universe using the Cosmic Microwave
  Background Polarization,''
  Astrophys.\ J.\  {\bf 583}, 24 (2003)
  [arXiv:astro-ph/0207591].
  
\bibitem{Hirata}
C.~M.~Hirata and U.~Seljak,
  ``Reconstruction of lensing from the cosmic microwave background
  polarization,''
  Phys.\ Rev.\  D {\bf 68}, 083002 (2003)
  [arXiv:astro-ph/0306354].
  
\bibitem{Holder2}
G.~P.~Holder, K.~M.~Nollett and A.~van Engelen,
  ``On Possible Variation in the Cosmological Baryon Fraction,''
  arXiv:0907.3919 [astro-ph.CO].
  
\end{thebibliography}
\end{document}